\documentclass[11pt]{article}
\usepackage{moriond,epsfig}

\usepackage[square,numbers,sort&compress,nonamebreak]{natbib}
\usepackage{hypernat}
\usepackage{color}
\usepackage{ifpdf}

\newif\ifcomment
\newif\ifprint
\printtrue

\ifpdf
\ifprint
 \usepackage[linkbordercolor={0 0 .8},%
             citebordercolor={.8 0 0},%
             urlbordercolor={.8 0 .8},%
             raiselinks=true,%
             pdfborder={0 0 1 [3]}]{hyperref}
\else
 \usepackage[pdftex,backref,pagebackref,colorlinks,
             linkcolor=blue,citecolor=blue,anchorcolor=blue,
             pagecolor=blue,urlcolor=red,a4paper]{hyperref}
\fi
\pdfoutput=1        
\pdfcatalog{/PageMode(/UseOutlines)} 
\else 
\ifprint
 \usepackage[linkbordercolor={0 0 .8},%
             citebordercolor={.8 0 0},%
             urlbordercolor={.8 0 .8},%
             raiselinks=true,%
             pdfborder={0 0 1 [3]}]{hyperref}
\else
 \usepackage[dvips,backref,pagebackref,colorlinks,
             linkcolor=blue,citecolor=blue,anchorcolor=blue,
             pagecolor=blue,urlcolor=red,a4paper]{hyperref}
\fi
\hypersetup{
  pdfauthor   = {Constantin Loizides},
  pdftitle    = {},
  pdfcreator  = {},
  pdfproducer = {},
  pdfsubject  = {},
  pdfkeywords = {}
 }
\fi

\graphicspath{{./pics/}}

\def\Journal#1#2#3#4{{#1} {\bf #2}, #3 (#4)}

\def\NIMA{{\em Nucl. Instrum. Methods} A}
\def\NPA{{\em Nucl. Phys.} A}

\def\PRL{\em Phys. Rev. Lett.}
\def\PRC{{\em Phys. Rev.} C}

\def\be{\begin{equation}}
\def\ee{\end{equation}}
\def\bea{\begin{eqnarray}}
\def\eea{\end{eqnarray}}

\newcommand {\snn}       {\sqrt{s_{\scriptscriptstyle{{\rm NN}}}}}

\newcommand {\dnde}      {\ensuremath{\mathrm{d}N/\mathrm{d}\eta}}

\newcommand {\av}[1]     {\ensuremath{\left< #1 \right>}}
\newcommand {\abs}[1]    {\ensuremath{\left| #1 \right|}}

\newcommand {\text}[1]   {$\mathrm{#1}$}
\newcommand {\hrefurl}[1]{\href{#1}{\url{#1}}}
\newcommand {\arxiv}[1]  {\href{http://www.arxiv.org/#1}{\mbox{arXiv:#1}}}
\newcommand {\Ref}[1]    {Ref.~\cite{#1}}
\newcommand {\Refs}[1]   {Refs.~\cite{#1}}

\newcommand {\fig}[1]    {fig.~\ref{#1}}

\newcommand {\hide}[1]   {\color{white}#1\color{black}}

\newcommand {\eg}    {e.g.}

\newcommand {\Ncoll} {\ensuremath{N_{\rm coll}}}
\newcommand {\Npart} {\ensuremath{N_{\rm part}}}

\newcommand {\gev}   {\mbox{${\rm GeV}$}}

\newcommand {\mom}   {\mbox{\rm GeV$\kern-0.15em /\kern-0.12em c$}}
\newcommand {\gmom}  {\mbox{\rm GeV$\kern-0.15em /\kern-0.12em c$}}
\newcommand {\mass}  {\mbox{\rm GeV$\kern-0.15em /\kern-0.12em c^2$}}
\newcommand {\mmass} {\mbox{\rm MeV$\kern-0.15em /\kern-0.12em c^2$}}
\newcommand {\mmom}  {\mbox{\rm MeV$\kern-0.15em /\kern-0.12em c$}}

\newcommand {\fm}    {\mbox{${\rm fm}$}}

\begin{document}
\vspace*{4cm}
\title{SCALING FEATURES OF SELECTED OBSERVABLES AT RHIC}

\author{Constantin LOIZIDES$^{4}$ for the PHOBOS collaboration\\
\vspace{2mm}
%
%
B.Alver$^4$,
B.B.Back$^1$,
M.D.Baker$^2$,
M.Ballintijn$^4$,
D.S.Barton$^2$,
R.R.Betts$^6$,
R.Bindel$^7$,
W.Busza$^4$,
Z.Chai$^2$,
V.Chetluru$^6$,
E.Garc\'{\i}a$^6$,
T.Gburek$^3$,
K.Gulbrandsen$^4$,
J.Hamblen$^8$,
I.Harnarine$^6$,
C.Henderson$^4$,
D.J.Hofman$^6$,
R.S.Hollis$^6$,
R.Ho\l y\'{n}ski$^3$,
B.Holzman$^2$,
A.Iordanova$^6$,
J.L.Kane$^4$,
P.Kulinich$^4$,
C.M.Kuo$^5$,
W.Li$^4$,
W.T.Lin$^5$,
S.Manly$^8$,
A.C.Mignerey$^7$,
R.Nouicer$^2$,
A.Olszewski$^3$,
R.Pak$^2$,
C.Reed$^4$,
E.Richardson$^7$,
C.Roland$^4$,
G.Roland$^4$,
J.Sagerer$^6$,
I.Sedykh$^2$,
C.E.Smith$^6$,
M.A.Stankiewicz$^2$,
P.Steinberg$^2$,
G.S.F.Stephans$^4$,
A.Sukhanov$^2$,
A.Szostak$^2$,
M.B.Tonjes$^7$,
A.Trzupek$^3$,
G.J.van~Nieuwenhuizen$^4$,
S.S.Vaurynovich$^4$,
R.Verdier$^4$,
G.I.Veres$^4$,
P.Walters$^8$,
E.Wenger$^4$,
D.Willhelm$^7$,
F.L.H.Wolfs$^8$,
B.Wosiek$^3$,
K.Wo\'{z}niak$^3$,
S.Wyngaardt$^2$,
B.Wys\l ouch$^4$\\
\vspace{3mm}
\small
%
%
%
%
$^1$~Argonne National Laboratory, Argonne, IL 60439-4843, USA\\
$^2$~Brookhaven National Laboratory, Upton, NY 11973-5000, USA\\
$^3$~Institute of Nuclear Physics PAN, Krak\'{o}w, Poland\\
$^4$~Massachusetts Institute of Technology, Cambridge, MA 02139-4307, USA\\
$^5$~National Central University, Chung-Li, Taiwan\\
$^6$~University of Illinois at Chicago, Chicago, IL 60607-7059, USA\\
$^7$~University of Maryland, College Park, MD 20742, USA\\
$^8$~University of Rochester, Rochester, NY 14627, USA\\}

\address{}

\maketitle

\abstracts{
We discuss several observables measured by PHOBOS that show 
common scaling features in Cu+Cu and Au+Au collisions at RHIC energies. 
In particular, we examine the centrality and energy dependence of the charged 
particle multiplicity, as well as the centrality dependence of the elliptic 
flow at mid-rapidity. The discrepancy between Cu+Cu and Au+Au of the final state 
azimuthal asymmetry~(elliptic flow), relative to the initial state geometry of 
the collision, can be resolved by accounting for fluctuations in the description
of the initial geometry. }

\section{Introduction}
The study of heavy-ion collisions at ultra-relativistic energies allows one
to combine experimental with theoretical efforts in the understanding of the 
phase-space properties of strongly-interacting matter. At extreme conditions
of high temperature and density~(compared to normal nuclear matter), numerical 
QCD calculations predict a phase transition to a system dominated by partonic, 
rather than hadronic, degrees of freedom. Indeed, one of the important 
conclusions from the discoveries at the Relativistic Heavy Ion Collider~(RHIC) 
is~\cite{phobos_whitepaper} that in Au+Au collisions an extremely dense, highly 
interacting system is formed, reaching energy densities much larger than 
$\sim1~\gev/\fm^{3}$, the characteristic scale for the QCD phase transition.

In this conference proceeding, we will concentrate on a few simple 
observations extracted from data collected by the PHOBOS experiment at RHIC.
The observed scaling rules address common features in heavy-ion collisions~(Cu+Cu, 
Au+Au) and allow for the comparison with simpler systems~(d+Au, p+p) 
in a broad range of collision energies~($\snn=19.6$ to $200~\gev$).
The presented data are collected by the multiplicity detector covering 
$\abs{\eta}<5.4$ and the near mid-rapidity magnetic spectrometer 
of the PHOBOS experiment~(described in detail in \Ref{phobos_nim}).

\begin{figure}[t]
\begin{minipage}[t]{70mm}
\includegraphics[trim= 0 -49 0 0, width=7cm]{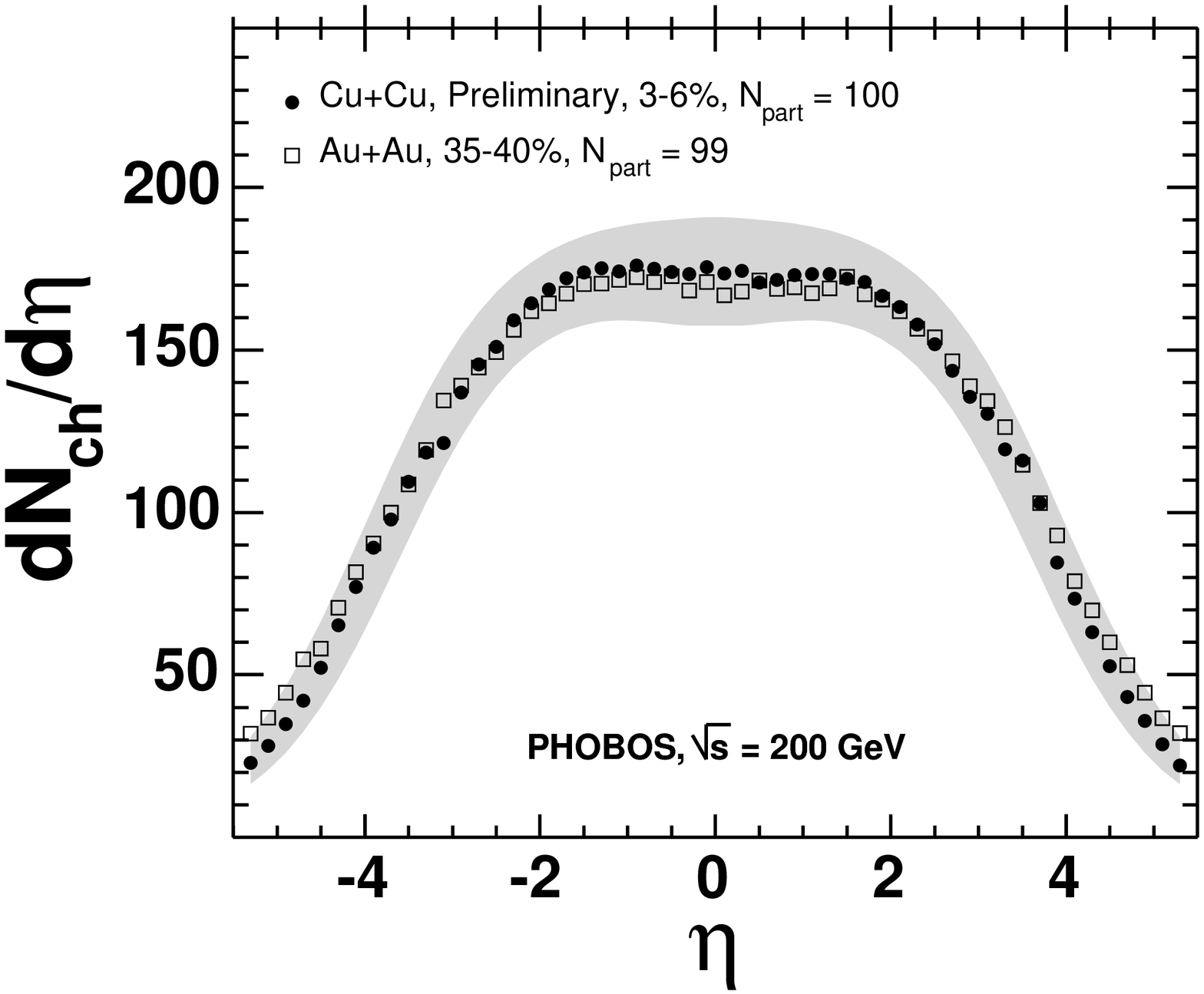}
\vspace{-1.0cm}
\caption{Pseudo-rapidity distribution for charged hadrons in Cu+Cu~(closed symbols) and Au+Au 
collisions~(open symbols) at $\sqrt{s_{_{NN}}} = 200~\gev$. The Cu+Cu and Au+Au centralities 
were selected to yield similar $\av{\Npart}$. The band indicates the systematic 
uncertainty for Cu+Cu~(90\%~C.L.). Errors for Au+Au are not shown.\hide{aaaaaaaaaaaaaa}}
\label{fig1}
\end{minipage}
\hspace{\fill}
\begin{minipage}[t]{80mm}
\includegraphics[width=8cm]{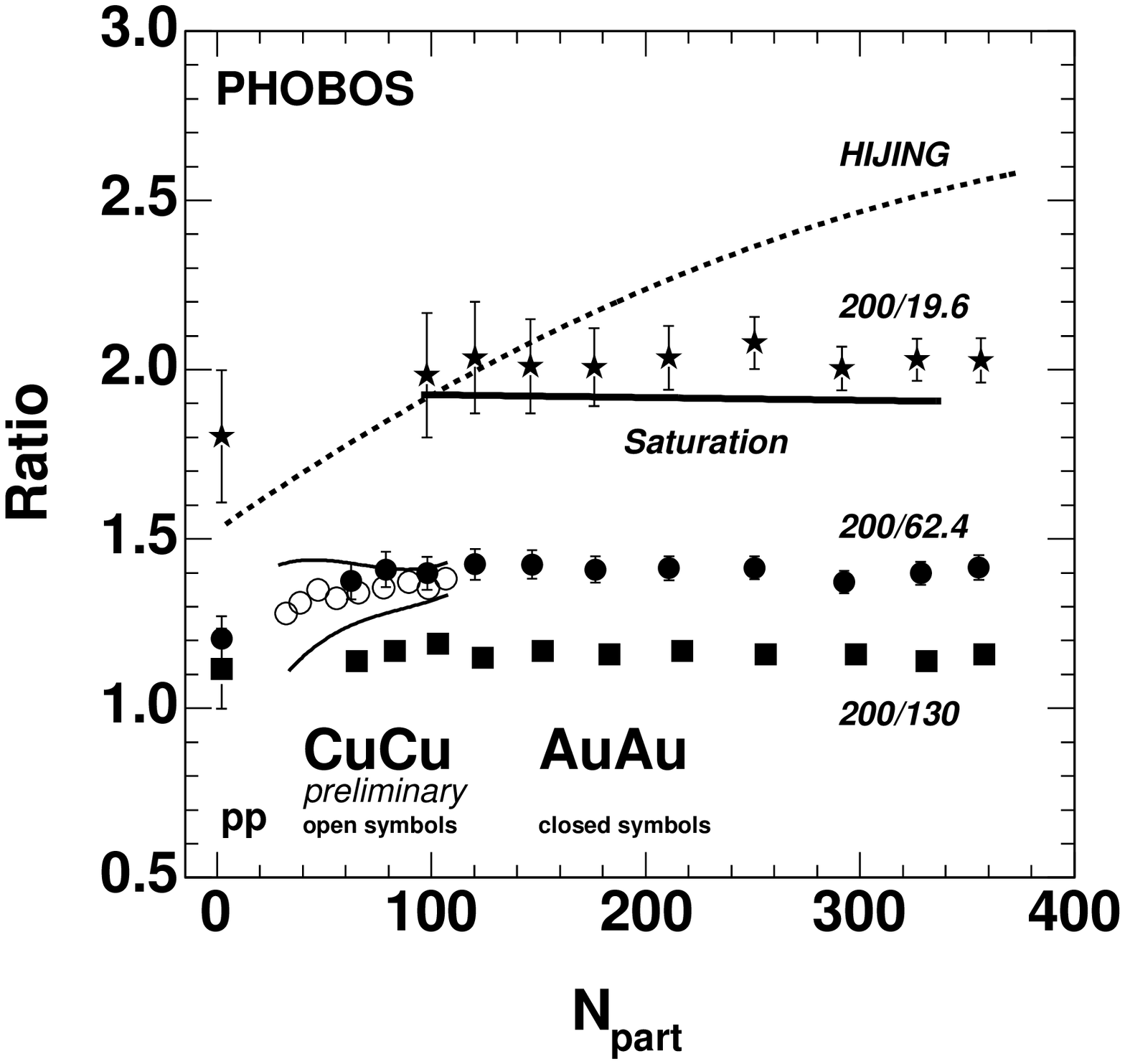}
\vspace{-1.0cm}
\caption{Ratio of mid-rapidity densities 
as a function of $\Npart$. Au+Au data are for energies 200/19.6, 200/62.4, 
200/130~(closed symbols) and Cu+Cu data for 200/62.4~(open symbols).
The two solid lines around the Cu+Cu ratio indicate additional uncertainty from 
the estimated trigger efficiency. 
Ratios for p+p are shown at 
\mbox{$\Npart=2$}. Model calculations are from \Refs{hijing,kharzeev}.\hide{aaaaa}}
\label{fig2}
\end{minipage}
\end{figure}

\section{Scaling features in heavy-ion data}
We focus on three topics of particle production in Cu+Cu and Au+Au collisions: 
the scaling of the overall charged-particle multiplicity, the factorization of the 
energy and centrality dependence of particle production and the connection between 
the observed final-state azimuthal particle distribution and the initial state 
geometry of the colliding system. 

\enlargethispage{0.5cm}
Before we can address these issues, it is important to quantify how particle 
production depends on the underlying geometry of the colliding species. The 
spatial overlap of the colliding nuclei, determined by the impact parameter, is 
described as the centrality of the collision. Centrality is typically parametrized 
by the number of participating nucleon pairs~($\Npart/2$), or the number of binary 
nucleon--nucleon collisions~($N_{coll}$), in the overlap region. 
Both quantities grow with increasing centrality~(decreasing 
impact parameter) of the collision and are calculated within a Monte-Carlo 
simulation of the Glauber model. They are finally related to data via the 
fractional cross-section of the nucleus--nucleus interaction from detailed 
comparison of measured versus simulated charged hadron multiplicity (for 
details see \Ref{phobos_qm05} and references therein). 

\subsection{System-size dependence of particle production}
In \fig{fig1} we show the charged hadron $\dnde$ distributions in Cu+Cu and Au+Au 
at $\snn=200~\gev$. Centrality bins are chosen such that the average number of 
participants in Cu+Cu is approximately equal to that in Au+Au.
This comparison, in combination with further studies for different centrality selections, 
leads to a simple scaling rule: If Cu+Cu and Au+Au collisions at the same collision 
energy are selected to have the same $\langle N_{part} \rangle$, the resulting 
charged hadron $\dnde$ distributions are nearly identical, both in the 
mid-rapidity particle density and the width of the distribution. The same is true
for $\dnde$ at $62.4~\gev$~(not shown, see \Ref{phobos_qm05}). Similar findings 
apply for transverse momentum distributions at both energies~\cite{phobos_cucuspec}.

\subsection{Factorization in energy and centrality}
In studying the centrality and energy dependence 
we have established a further, more subtle, scaling relationship 
that holds for all of the above mentioned observables.
As first described in~\Ref{phobos_tracklets} for the mid-rapidity density,
the increase in particle production per participant with increasing $\Npart$ is 
independent of collision energy over the full energy range of RHIC from $19.6$ to 
$200~\gev$.
\enlargethispage{0.5cm}
This is illustrated in \fig{fig2}, where we show the ratio of mid-rapidity 
densities as a function of $\Npart$ relative to the $200~\gev$ data. All ratios, 
from $200/130$ to $200/19.6$, are flat within the experimental uncertainty.
The factorization is obviously violated in a model, such as HIJING~\cite{hijing}, 
that determines overall particle production using a superposition of independent 
``soft'' contributions scaled with $\Npart$ and ``hard'' contributions scaled with 
$\Ncoll$. However, the observed factorization is fulfilled in approaches based 
on the ideas of parton saturation, as \eg~in the calculation from Kharzeev et 
al.\ for the 200/19.6 ratio~\cite{kharzeev}. See \Ref{phobos_qm05} for a more 
complete overview on factorization.

\subsection{System-size dependence of elliptic flow}
\begin{figure}[t]
\includegraphics[width=\textwidth]{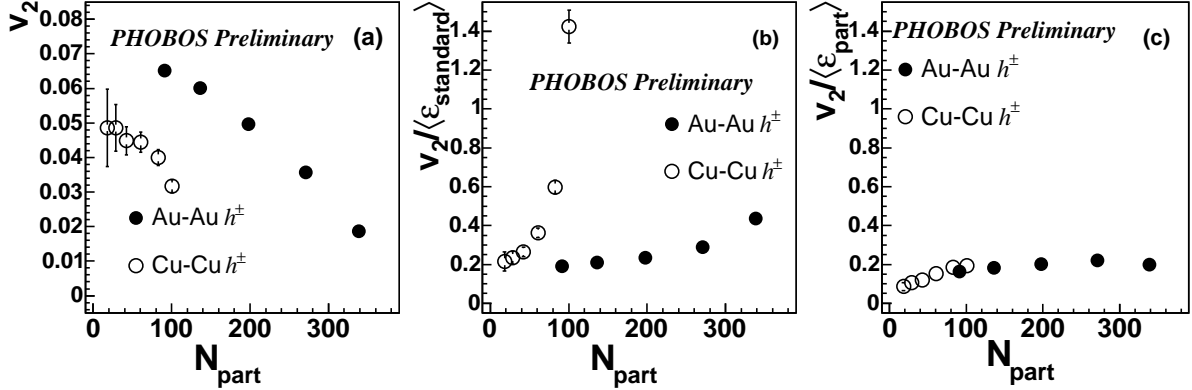}
\caption{Unscaled and scaled elliptic flow as a function of $\Npart$ at mid-rapidity
($\abs{\eta}<1$) for Au+Au and Cu+Cu at $\snn=200~\gev$:
(a) unscaled $v_2$, 
(b) $v_{2}$/$\langle \varepsilon_{\rm standard} \rangle$, 
(c) $v_{2}$/$\langle \varepsilon_{\rm part} \rangle$.
Only statistical errors are shown.}
\label{fig3}
\end{figure}

The elliptic flow at mid-rapidity, $v_2$, as obtained from the angular distribution of 
particles wrt.~the reaction plane, is shown in \fig{fig3}(a) as a function 
of $\Npart$ for Cu+Cu~\cite{phobos_qm05} and Au+Au~\cite{phobos_v2} collisions at 
$200~\gev$.
Its measurement provides important constraints on the hydrodynamical evolution of the 
collision dynamics and gives insight into the connection between initial state and 
final state effects.
The initial geometric asymmetry of the collision is typically assumed to be given by the 
eccentricity of the overlap region of the two nuclei. 
The average eccentricity for a certain centrality class is obtained from a Glauber 
simulation. The eccentricity, $\varepsilon_{\rm standard}$, of the distribution 
of participating nucleons relative to the reaction plane is commonly calculated by fixing 
the minor axis of the overlap ellipse to be along the impact parameter vector.
Since the eccentricity for a fixed value of $\Npart$ depends on the size of the colliding 
species, we scale out the difference in the initial geometry and compare 
$v_{2}$/$\langle \varepsilon_{\rm standard} \rangle$ 
in \fig{fig3}(b) for the two systems. 
The scaled flow in Au+Au is quite flat as a function of centrality showing that the 
flow in Au+Au relatively closely follows the initial state eccentricity.
However, comparing the scaled Cu+Cu data to the Au+Au data leads to the paradoxical 
conclusion that for the same $\Npart$, the smaller system (Cu+Cu) is more effective 
in translating the initial eccentricity into the final state anisotropy. 
A possible resolution to this paradox that produces consistent results in Cu+Cu and 
Au+Au, as shown in \fig{fig3}(c), can be obtained by refining the definition of the 
initial state eccentricity.
Our alternative measure of the eccentricity, $\varepsilon_{\rm part}$,
is based on the observation that $\varepsilon_{\rm standard}$ underestimates fluctuations 
in the actual participants distribution, which would generally lead to too small 
of a mean value. 
We define $\varepsilon_{\rm part}$ in each centrality bin by calculating the eccentricity for 
each Glauber event relative to the principal axes of the actual participant distribution~(see 
\Ref{manlyqm05}). 
By construction, $\varepsilon_{\rm part}$ is positive definite and leads to a finite average 
value even for the most central events. 
The smaller number of colliding nucleons makes the difference between $\varepsilon_{\rm standard}$ 
and $\varepsilon_{\rm part}$ particularly important for the Cu+Cu system. 

\section{Summary}
\enlargethispage{0.5cm}
We address selected simple scaling observations extracted from Cu+Cu and Au+Au data collected 
by the PHOBOS experiment at RHIC. We find that particle production per participant nucleon is 
very similar in Cu+Cu and Au+Au collisions for collision centralities with equivalent number 
of participants. The Au+Au data at RHIC, and to a similar extent also the Cu+Cu data, exhibit
the factorization of energy and centrality dependence in particle production at mid-rapidity. 
The elliptic flow in Cu+Cu at mid-rapidity is surprisingly large relative to the expected 
average initial state anisotropy, especially when estimated via the ``standard eccentricity''. 
The result can be quantitatively understood, if one accounts for fluctuations in the initial 
collision geometry, which leads to scaling of $v_2$ relative to the ``participant eccentricity'' in 
both systems.

\section*{Acknowledgments}
%
%
%
%
The present work was partially supported by U.S. DOE grants 
DE-AC02-98CH10886,
DE-FG02-93ER40802, 
DE-FC02-94ER40818,  
DE-FG02-94ER40865, 
DE-FG02-99ER41099, and
W-31-109-ENG-38, by U.S. 
NSF grants 9603486, 
and 0245011,        
by the Polish KBN grant 1-P03B-062-27~(2004-2007),
by the NSC of Taiwan Contract NSC~89-2112-M-008-024, and
by the Hungarian OTKA grant (F~049823).


\end{document}